\documentclass[12pt,preprint]{aastex}
\usepackage{amsmath}
\usepackage{amssymb}

\newcommand{\vect}[1]{\mathbf{#1}}

\newcommand{\ls}[1]{{\mathcal{L}_{#1}}}
\newcommand{\rf}{{\mathrm{ref}}}
\newcommand{\conv}{{\mathrm{conv}}}
\newcommand{\surf}{{\mathrm{surf}}}
\newcommand{\xray}{{\mathrm{X}}}
\newcommand{\mg}{\mbox{\ion{Mg}{2}}}

\shorttitle{Dynamos and X-ray Flux in Stars}
\shortauthors{Bercik et al.}

\begin{document}
\title{Convective Dynamos and the Minimum X-ray Flux in Main Sequence Stars}
\author{D.~J.~Bercik$\,^{1}$, G.~H.~Fisher$\,^{1}$, 
   Christopher M.~Johns-Krull$\,^{2}$ and W.~P.~Abbett$\,^{1}$}
\affil{$\,^{1}$ Space Sciences Laboratory, University of California,
   Berkeley, CA 94720-7450}
\affil{$\,^{2}$ Department of Physics and Astronomy, Rice University,
   6100 Main St., MS-108, Houston, TX 77005}

\begin{abstract}
The objective of this paper is to investigate whether a convective
dynamo can account quantitatively for the observed lower
limit of X-ray surface flux in solar-type main sequence stars.  Our approach 
is to use 3D numerical simulations of a turbulent dynamo driven by convection 
to characterize the dynamic behavior, magnetic field strengths, and filling 
factors in a non-rotating stratified medium, and to predict these magnetic 
properties at the surface of cool stars.  
We use simple applications of stellar structure theory for the convective 
envelopes of main-sequence stars to scale our simulations to the outer 
layers of stars in the F0--M0 spectral range, which allows us to estimate 
the unsigned magnetic flux on the surface of non-rotating reference 
stars.  With these estimates we use the recent results of 
\citet{Pevtsov03} to predict the level of X-ray emission from such a 
turbulent dynamo, and find that our results compare well with 
observed lower limits of surface X-ray flux.   If we scale our
predicted X-ray fluxes to \ion{Mg}{2} fluxes we also find good agreement 
with the observed lower limit of chromospheric emission in K dwarfs.
This suggests that dynamo action from a convecting, non-rotating plasma
is a viable alternative to acoustic heating models as an explanation for the 
basal emission level seen in chromospheric, transition region, and coronal 
diagnostics from late-type stars.
\end{abstract}
\section{Introduction}\label{sec:intro}
Understanding the origin of magnetic activity in stars has been an 
important research area in astronomy and astrophysics for many years.  
Imaged solar observations show a clear link between magnetic fields 
and the formation of heated plasma in the chromosphere, transition 
region and corona, especially in active regions (see \citealt{Fisher98}).
Active regions are believed to form from loops of magnetic
flux that emerge from the base of the solar convection zone (see e.g., 
reviews by \citealt{Fisher00,Fan04}
\footnote{\url{http://www.livingreviews.org/lrsp-2004-1}}). The 
large-scale magnetic field on the Sun is believed to originate via a 
global-scale dynamo, powered mainly by the velocity shear (differential 
rotation) between the convection zone and the radiative interior.  It is 
often assumed that the amount of differential rotation increases as 
the rotation rate itself increases.  For example, theoretical studies of 
differential rotation mechanisms such as the ``$\Lambda$ 
effect'' \citep{Rudiger89} find that the amount of differential rotation 
generated by Reynolds stresses is proportional to the rotation rate.

\citet{Skumanich72} was among the first to propose an observational 
connection between rotation rate and the level of activity in stars 
through dynamo action.  \citet{Noyes84} went on to show a clear 
correlation between the \ion{Ca}{2} H+K surface flux and rotation 
rate, arguing that the correlation is made better with the inclusion 
of a convective turnover time via the Rossby number.  \citet{Vilhu84} 
finds a similar result for chromospheric, transition region, and
coronal emission diagnostics.  \citet{Basri87} shows that the activity of 
dwarf and RS CVn stars both correlate well with rotation rate, but 
with a slight offset between the two classes of stars, while the two 
classes appear to follow the same relationship when the rotation period 
is divided by the convective turnover time (see also 
\citealt{Johns-Krull00}).  These studies provide strong observational 
support for the existence of the so-called $\alpha$-$\Omega$ dynamo 
(e.g., \citealt{Durney78}), which relies on the interaction of 
convection and rotation --- specifically, differential rotation (the 
``$\Omega$'' effect) and field regeneration by cyclonic motions (the 
``$\alpha$'' effect) --- to generate a magnetic field in a fashion 
similar to that believed responsible for the large-scale solar dynamo.  

For cool stars, a straightforward application of the relationship 
between stellar rotation and magnetic activity described above leads 
one to conclude that those stars that rotate slowly or do not rotate 
(hereafter collectively termed ``non-rotating stars'') should show 
little or no activity diagnostics.  However, this is not the case.
There appears to be a lower limit to the emission of activity 
indicators in the chromosphere, transition region, and corona. 
Given the expected absence of a global dynamo in non-rotating stars,
this heating is usually attributed to a ``basal'' non-magnetic
mechanism such as heating by acoustic waves.
\citet{Schrijver87} first introduced the term 
``basal flux density'' to refer to that portion of the chromospheric 
and transition region line emission which results from a process other 
than the solar-like cyclic dynamo activity (termed ``excess'' emission). 

Evidence for a basal flux and the contention that it is acoustic in
origin has been based on three observational arguments.  First, plots 
of the \ion{Ca}{2} H+K chromospheric flux (as observed in the 
Mt.~Wilson HK project, \citealt{Vaughan78}) versus stellar color show 
a lower boundary up against which the stellar observations appear to 
cluster \citep{Rutten84,Schrijver87,Rutten91}.  Such a lower boundary 
is also apparent in other diagnostics such as \ion{Mg}{2} h and k and 
\ion{Si}{2} \citep{Schrijver87} as well as in \ion{C}{2}, \ion{C}{4}, 
and \ion{Si}{4} \citep{Rutten91}.  A second line of analysis examines 
flux-flux diagrams for samples of stars and seeks to make these as 
tight as possible by subtracting a color dependent basal emission 
level from one or more of the diagnostics under consideration 
\citep{Mewe81,Schrijver83,Schrijver87,Schrijver89,Rutten91}.
The basal emission level is suggested to be acoustic in origin, 
while the ``excess'' above this level is believed to be magnetic in 
origin.  In cases where a basal flux is determined from these 
flux-flux studies, its value is similar to the lower limit of the 
flux found in flux-color diagrams.  A third argument for the 
existence of a basal flux comes from studies connecting rotation and 
activity.  \citet{Schrijver89} extrapolate the color dependent 
rotation-activity relationship in \ion{Ca}{2} H+K to zero rotation 
velocity, finding a non-zero activity level which they 
identify as the basal flux level.  This procedure produces a color 
dependent basal emission similar in strength to that found from 
flux-color and flux-flux analyses, and this basal emission is again 
argued to be non-magnetic in origin.

Theoretical estimates of the acoustic heating of chromospheres have been
in the literature for some time (see review by \citealt{Schrijver95}).  
We mention here only some of the more recent results which now show good
agreement with the observationally inferred basal emission levels.
\citet{Buchholz98} compute time dependent models of basal heating
for monochromatic acoustic wave models, solving the hydrodynamic equations
together with the radiative transfer and statistical equilibrium equations 
in order to predict basal emission levels in both the \ion{Ca}{2} H+K
and \ion{Mg}{2} h+k diagnostics.  These models produce the observed
dependence of the lower bound of emission with stellar color, though the
magnitude of the emission depends rather sensitively on whether complete
or partial redistribution is assumed for the line transfer.  Unfortunately,
partial redistribution works best for \ion{Mg}{2} while complete 
redistribution gives a closer match to the \ion{Ca}{2} observations.  
\citet{Mullan94a,Mullan94b} extend their acoustic wave models to lower 
mass column densities and find the formation of a relatively cool corona 
($\sim 0.6$ MK) when applied to F stars and to M dwarfs ($T=0.7$--$1.0$ MK).  
While this temperature for M dwarfs is lower than the coronal temperatures 
found observationally for both dMe and dM stars \citep{Giampapa96}, 
\citet{Mullan94a} do find that their acoustically heated models can
produce a surface flux in excess of that observed on the least X-ray
active dM stars.  Given the relative success of wave propagation models
at predicting the basal level of chromospheric (and possibly coronal)
emission, \citet{Judge98} examined high spectral resolution 
observations of \ion{C}{2} lines in basal level stars and in quiet regions 
on the Sun to look for evidence of upward propagating shock waves as
predicted by acoustic heating models.  No such evidence was found,
prompting \citet{Judge98} to call into question the non-magnetic
origin for basal emission.

The role of basal flux in the observed X-ray emission is 
more uncertain observationally compared to chromospheric and transition 
region diagnostics.  The early flux-flux studies 
\citep{Mewe81,Schrijver83,Schrijver87,Schrijver89} did not attempt to find 
a basal level in soft X-ray emission.  \citet{Rutten91} 
did find a strongly color dependent lower bound on the X-ray emission of 
stars in his sample, but this was likely due to instrumental detection 
limits, as the recent volume limited study by \citet{Schmitt97} clearly 
shows that the lower bound on X-ray emission in cool stars has a very weak 
color dependence at best; which, if it exists, is in the opposite sense to 
that found by \citet{Rutten91}.  While their X-ray data probably suffered 
from detection limitations, it should be noted that \citet{Rutten91} do not 
find a basal level to the X-ray emission --- they find that all of the 
X-ray emission is ``excess'' in nature, and therefore likely magnetic in 
origin.  In a later study, \citet{Mullan94b} argued that the 
observed X-ray emission from late A and early F dwarfs is basal (and 
therefore acoustic) in nature, while \citet{Mullan94a} and \citet{Mullan96} 
have argued that 90\% or more of the X-ray emission from dM stars is basal 
(acoustic) in origin.

In the more recent, volume-limited study of nearby stars, \citet{Schmitt97} 
finds that essentially all stars with outer convection zones emit X-rays 
with characteristics similar to the solar coronal emission.  It seems, 
therefore, that all ``cool stars'' possess coronae.  Do they then also 
possess magnetic fields?  \citet{Schmitt97} notes that for the nearby 
cool stars, there is a well defined minimum X-ray surface flux of 
$\log F_{\xray} \sim 3.7$ (in cgs units) which appears largely independent of 
stellar color.  Such a lower bound to the X-ray emission is reminiscent 
of the idea of ``basal'' surface flux in chromospheric and transition 
region emission in cool stars, which is usually assumed to be acoustic in 
origin, as noted above.  Recent work analyzing both solar and stellar data 
\citep{Pevtsov03} indicates a clear and unambiguous relationship between 
unsigned magnetic flux and coronal X-ray emission that extends over twelve 
orders of magnitude.  In that paper, observations ranging from small patches
of the quiet Sun to the most active pre-main sequence stars were analyzed.  
At the small-scale end of this study, Kitt Peak magnetograms and Yohkoh SXT 
images were averaged over a $4$ heliographic degree square area centered 
at the equator and central meridian of the Sun, with reported magnetic 
fluxes and X-ray fluences normalized to a single arc-second square area on 
the solar disk, corresponding to the approximate size of a single Kitt Peak
magnetogram pixel \citep{Pevtsov04}.  This study lends further credence to 
the idea that at least for X-ray emission, a ``basal'' heating mechanism 
may indeed be magnetic in origin, but unlike the ``excess'' emission from 
global fields, the magnetic flux has little or no connection to stellar 
rotation.

High resolution observations of the Sun indicate a viable magnetic 
origin for this ``basal'' magnetic activity emission:  The existence of a 
small-scale magnetic component which appears to exist independent of the 
large-scale fields that define the solar cycle, and which had been predicted 
to be generated locally by turbulent convective motions 
\citep{Meneguzzi86,Durney93}.  Detailed studies of the small-scale magnetic 
field on the Sun from the MDI instrument on SOHO \citep{Title98,Hagenaar03} 
find strong evidence that small-scale mixed-polarity magnetic concentrations 
observed in the quiet Sun show only a weak solar-cycle dependence.  By 
studying the detailed evolution of this field over time scales of 
roughly one day, the authors conclude that this flux is formed by the 
emergence of small-scale bipoles (e.g., ephemeral active regions, 
\citealt{Harvey93}) and dissipated by flux cancellation, where the 
collision of two opposite polarities appears to result in their mutual 
disappearance \citep{Simon01,Parnell02}.  \citet{Title98} estimate a flux 
replenishment time scale of forty hours, while the more recent study of 
\citet{Hagenaar03} estimate a replenishment time scale of 
sixteen hours.  In any case, the quiet Sun magnetic flux --- however it 
is generated ---  appears to correctly predict the X-ray fluence for the 
quiet Sun in the study of \citet{Pevtsov03}.  More recent observations 
\citep{Lin99,Khomenko03,Bueno04} using spectropolarimetric instruments in 
conditions of extremely good seeing have found evidence of a mixed 
polarity magnetic field on even smaller, sub-granulation spatial scales.  
Taken together, these measurements imply that turbulent convection may 
produce a small-scale mixed polarity field over a very wide range of spatial 
scales in the Sun.

Complementing the observational evidence for small-scale magnetic fields is
a broad base of theoretical work.  Research in the field of ``fast dynamos'' 
(e.g., \citealt{Vainshtein96,Tanner03}) indicates that in the presence of 
wide classes of chaotic flow fields, magnetic energy can grow very 
efficiently, even in the absence of rotation.  In these studies, 
the magnetic induction equation is solved kinematically in the presence of a
prescribed flow field.  If the fluid streamlines are sufficiently chaotic,
the growth of magnetic energy can proceed exponentially quickly.
Much of the research in this area focuses on highly idealized 
flow fields (e.g. \citealt{Vainshtein96}) and does not 
account for the role of the Lorentz force in limiting the growth of
the magnetic field.

The major theoretical breakthrough in this area was made by 
\citet{Cattaneo99}, who demonstrated via a 3D MHD simulation in the 
Boussinesq approximation that a small seed field embedded in a turbulently 
convecting, highly conducting plasma can indeed grow exponentially until 
the magnetic energy reaches a level of $10$--$20$\% of the kinetic energy 
of the convective motions.  Subsequent work by \citet{Thelen00} has shown how 
variation of boundary conditions and fluid mechanics parameters affect the 
development of the dynamo, while \citet{Emonet01} have focused on computing 
observational diagnostics of the small-scale disordered field.

Turbulent convective dynamo models of this type must not be confused with 
global-scale dynamo models driven by kinetic helicity in small-scale turbulent 
motions,
which require rotation to generate an $\alpha$ effect.  An example
of such a model is that of \citet{Kuker99}, who used an $\alpha^2$ model 
to understand magnetic field generation in rotating, fully-convective 
naked T-Tauri stars, which are believed to have little or no differential 
rotation \citep{Johns-Krull96}.  The idea is that helical
turbulence driven by Coriolis forces can impart a systematic twist to field
lines encountering the convective eddies, leading eventually
to a large-scale steady magnetic field.  In contrast, the turbulent dynamo
that we believe to be responsible for the small-scale disordered field
in the quiet Sun is unaffected by rotation, since the convective turnover
time in the topmost layers of the solar convection zone is far smaller than
the solar rotation period.

The growing evidence for the importance of small-scale magnetic fields in 
stellar atmospheres and the observational evidence for a limit to the X-ray 
emission in stellar coronae leads to the question:  Is the turbulent convective
dynamo a viable mechanism to generate enough X-ray flux to account for the 
observational limit?  To address this question, in \S~\ref{sec:description}  
we extend the investigation of field generation by a turbulent dynamo beyond 
the Boussinesq approximation 
into the ``anelastic'' regime \citep{Gough69,Lantz99,Fan99} 
--- an approximation to the 3D MHD equations that allows 
for gravitational stratification while still filtering out acoustic waves.  
Here, we describe the numerical model and present a detailed 
analysis of our dynamo simulations.  In \S~\ref{sec:connect}, we describe 
our method of utilizing the simulation results to predict the level of 
X-ray emission in main sequence stars, in \S~\ref{sec:results} we
present our results, and finally in \S~\ref{sec:discussion} we discuss
the implications of our results and summarize our findings.

\section{Description of the Model Dynamo} \label{sec:description}
We solve the 3D system of MHD equations 
in the anelastic limit (see \citealt{Lantz99, Fan99} and 
references therein for a description of the anelastic formalism and 
a discussion of the techniques employed to numerically solve the 
anelastic system).  Briefly, the anelastic approximation results 
from a scaled variable expansion of the fully-compressible MHD
equations.  Density fluctuations in the continuity equation
are of high order in the expansion and are neglected --- this 
effectively filters out rapidly propagating acoustic waves, 
and provides a significant computational savings.  The density 
fluctuations are retained, however, in the buoyancy force term of 
the momentum equation, which allows for stratification within the
computational domain.  Thus, the anelastic formulation
is an intermediary approximation between a fully-compressible 
treatment (see e.g., \citealt{Tobias2001,Dorch2004}) and a Boussinesq 
model (see e.g., \citealt{Cattaneo99}).  This treatment is well-suited 
for describing the high-$\beta$ plasma of stellar interiors, 
but cannot directly model surface plasmas where 
$\beta\approx 1$ and the acoustic Mach number becomes large.

In particular, we solve a non-dimensional form of the anelastic 
MHD equations in a horizontally periodic, vertically closed 
Cartesian domain.  Our non-rotating rectangular box spans 
five pressure scale heights vertically, which corresponds to a 
density difference of $\sim 20$ between the upper and lower boundaries.  
The resolution of the Cartesian domain is $288\times288\times72$, 
which gives an aspect ratio for the box of $4:4:1$.  The 
non-dimensional parameters $R_e$ and $P_r$ (the Reynolds
number and Prandtl number, respectively) are defined
as $R_e\equiv \rho_{\rf}v_{\conv}H_{\rf}/\mu$ and $P_r\equiv\mu/\kappa$, 
and are set to $750$ and unity respectively (the Reynolds number 
was chosen to be as large as possible without introducing numerical
artifacts).  Here, $H_{\rf}$ denotes the pressure scale height at
the base of the domain, $\kappa$ refers to the coefficient of thermal
conductivity, and $\mu$ is the coefficient of dynamic viscosity
(assumed constant).  The convective velocity is measured in units of 
$v_{\conv}$, where $v_{\conv}\equiv (\delta_{\rf}gH_{\rf})^{1/2}$, 
$\delta_{\rf}$ is the non-dimensional super-adiabaticity, and $g$ is the 
constant vertical gravitational acceleration.

\begin{figure}
\epsscale{1.00}
\plotone{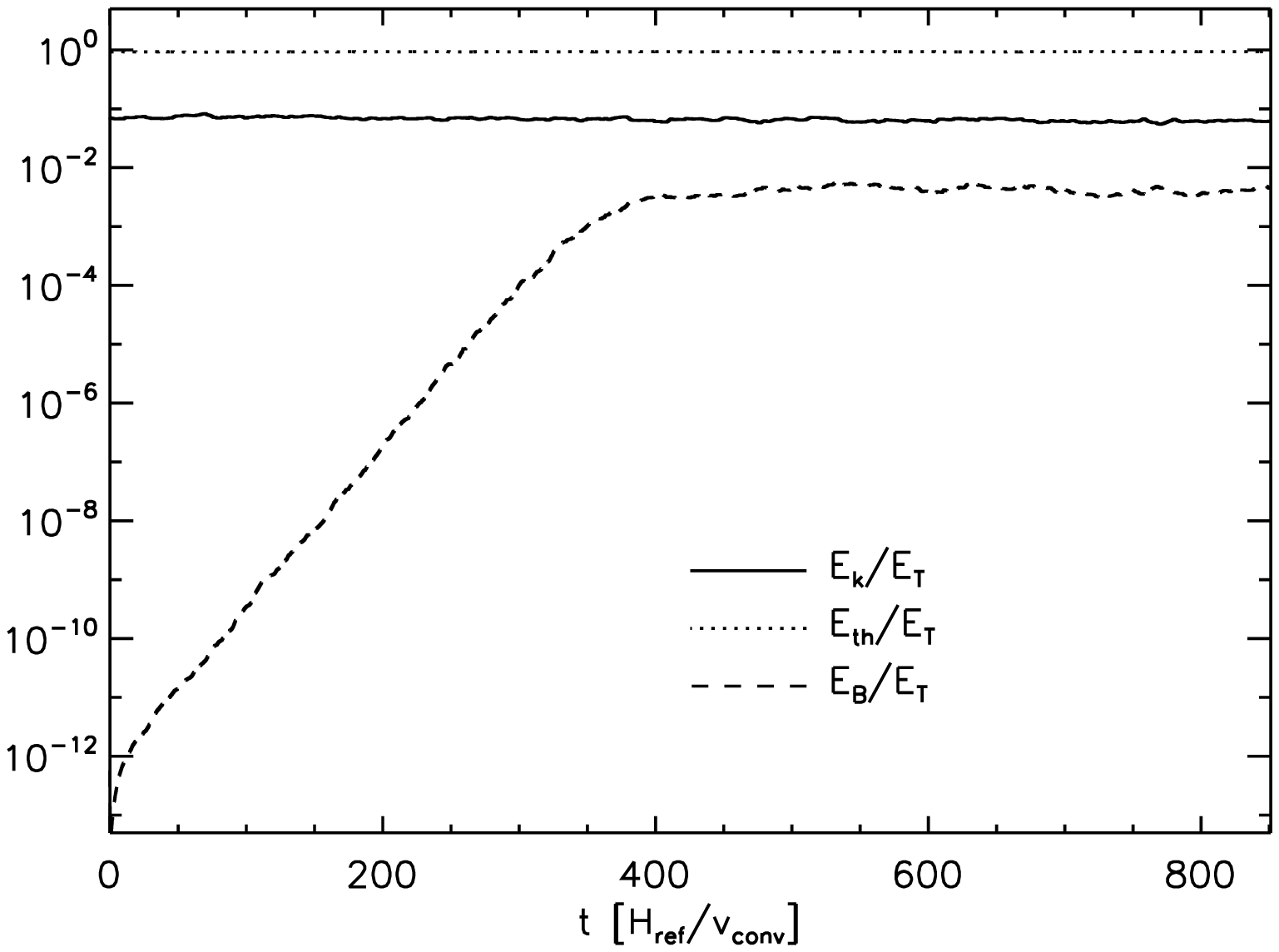}
\caption{Shown are the kinetic (solid line), thermal (dotted line), and
magnetic (dashed line) energy fluctuations (normalized by the sum of the 
three) as a function of time (in units of $H_{\rf}/v_{\conv}$).  
\label{fig:log_energy_relaxation}}
\end{figure}

\begin{figure}
\epsscale{1.00}
\plotone{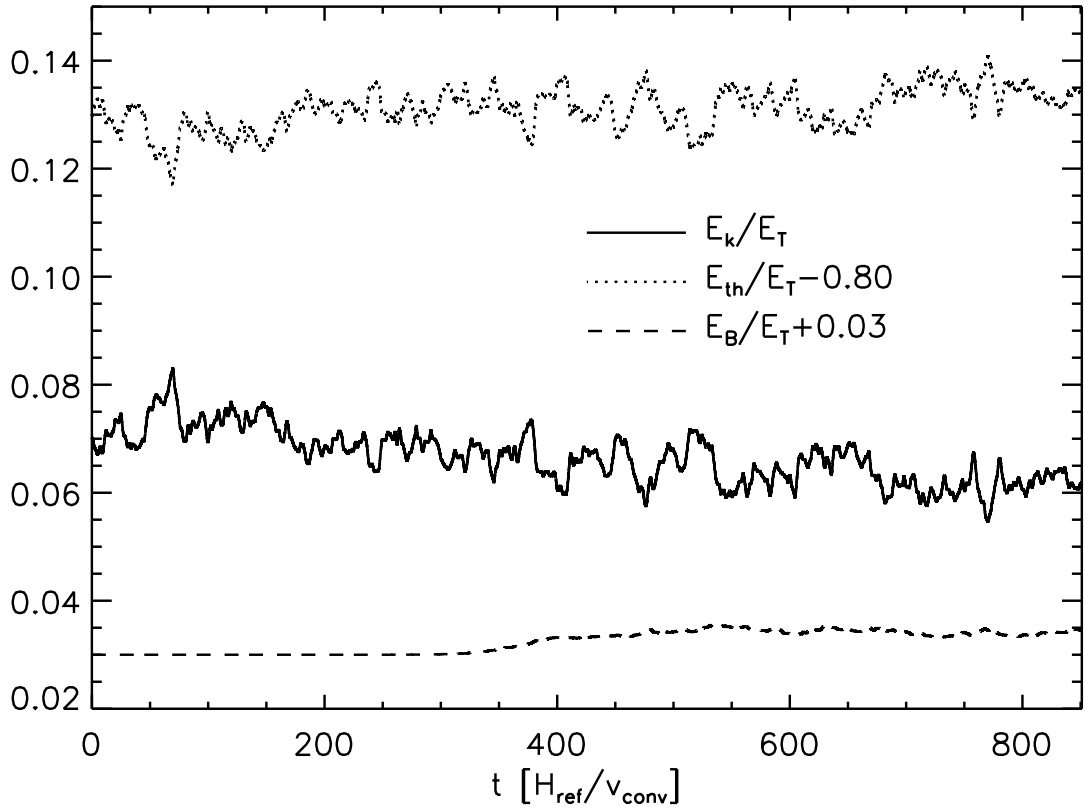}
\caption{Same as Figure~\ref{fig:log_energy_relaxation} except
on a linear scale.  To facilitate comparison, the magnetic and
thermal energies are vertically shifted. 
\label{fig:energy_relaxation}}
\end{figure}

We begin by dynamically and thermally relaxing a purely
hydrodynamic model convection zone.  We initiate convection
by introducing a small random entropy perturbation within
a computational domain with a prescribed background entropy 
gradient (see \citealt{Fan03,Abbett04}), and allow the simulation to 
progress past the time when the convective velocity field 
attains a statistically steady state.  We then set the 
magnetic Reynolds number $R_m\equiv v_{\conv}H_{\rf}/\eta$ of the
simulation to $1000$, and introduce a small, dynamically 
unimportant magnetic seed field (after $t=164\,H_{\rf}/v_{\conv}$ of
the field-free relaxation run). The seed field self-consistently 
grows and evolves within the computational domain as the run
progresses (see Figures~\ref{fig:log_energy_relaxation} and 
\ref{fig:energy_relaxation}).  
The vertical boundary conditions on the magnetic field are 
stress-free and non-penetrating at the bottom, and the field at the 
upper boundary is assumed potential.  We note that unlike the simulations 
of \citet{Abbett04}, our potential field upper boundary condition does 
allow for a small amount of horizontally-directed signed magnetic 
flux to diffuse out of the domain.  Also note that our ``global'' 
diffusion timescale of $t_D=1000\,H_{\rf}/v_{\conv}$ --- the characteristic 
time over which a magnetic structure diffuses across $H_{\rf}$ (as specified 
by our choice of magnetic Reynolds number) --- greatly exceeds the 
characteristic convective timescale of the simulation.

\begin{figure}
\epsscale{.65}
\plotone{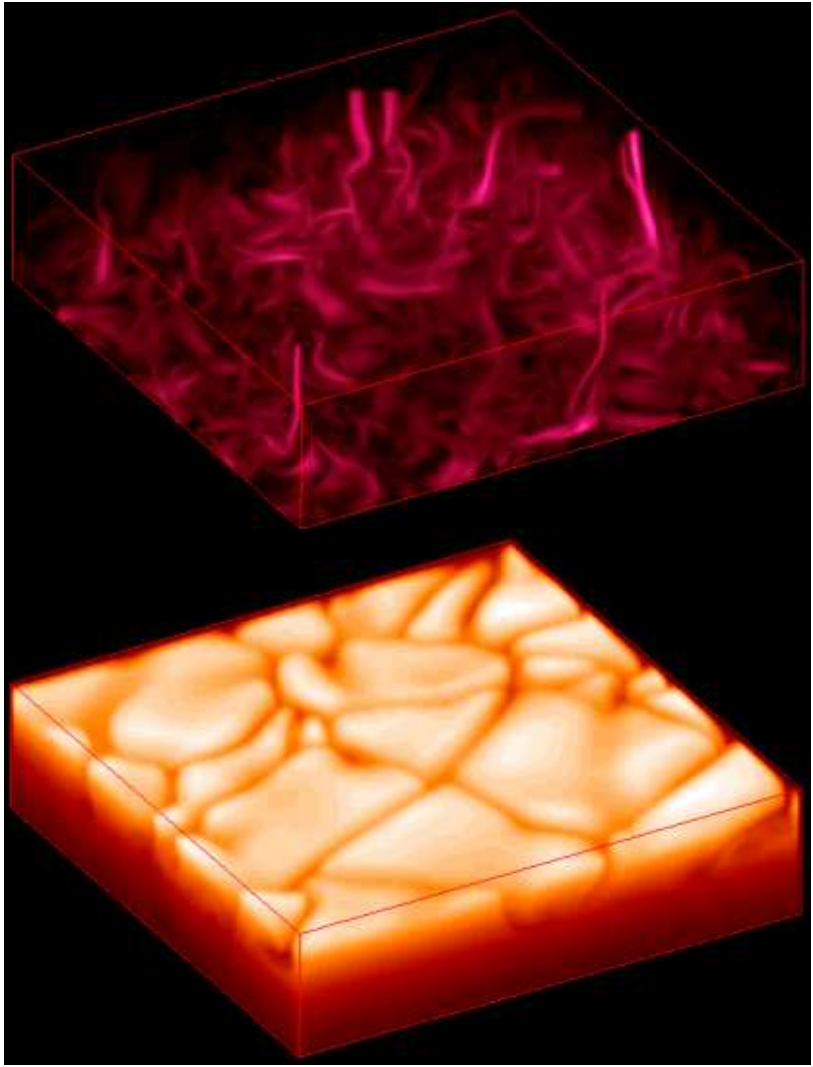}
\caption{Volume rendering of the magnetic field strength $|\vect{B}|$
(upper frame) and the corresponding entropy perturbations (lower frame)
at $t=710\,H_{\rf}/v_{\conv}$ (well after saturation).  
\label{fig:volume}}
\end{figure}

\begin{figure}
\epsscale{.90}
\plotone{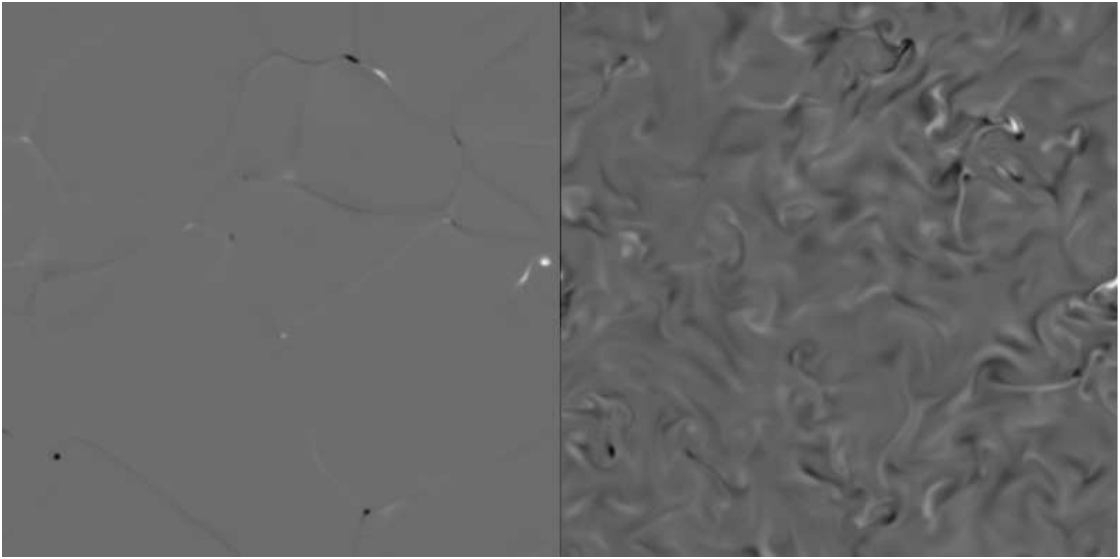}
\caption{The vertical component of the magnetic field 
at $t=710\,H_{\rf}/v_{\conv}$ along a horizontal slice taken 
at $z_u=2.06\,H_{\rf}$ near the top of the simulation domain 
(left frame), and at and $z_l=0.45\,H_{\rf}$ near the bottom (right 
frame).  
\label{fig:mgrams}}
\end{figure}

Figure~\ref{fig:log_energy_relaxation} shows the temporal evolution
of the total kinetic ($E_k$), thermal ($E_{th}$), and magnetic ($E_B$)
energy fluctuations (each integrated over the entire domain).  
The quantities are 
normalized by the sum of the three ($E_T\equiv E_k+E_{th}+E_B$).
We find that the energy of the seed field (initially at $10^{-12}$
of the total kinetic energy) increases exponentially with a 
growth time of $\sim 16\,H_{\rf}/v_{\conv}$ until the
magnetic field becomes dynamically important 
($t\approx 350\,H_{\rf}/v_{\conv}$), after which the increase becomes 
approximately linear in time.  The magnetic energy fully saturates to a 
time-averaged value of $6.7$\% of the total kinetic energy after 
$t\approx 600\,H_{\rf}/v_{\conv}$, and ranges from $5.5$--$8.5$\% of $E_k$. 
Figure~\ref{fig:energy_relaxation} shows the energies on a linear
scale --- as the magnetic energy increases, the kinetic
energy decreases, and the thermal energy fluctuations of the plasma
increase.  Figure~\ref{fig:volume} is a volume rendering of both 
the entropy perturbations (bottom frame) and magnetic 
field strength $|\vect{B}|$ well after the field has saturated 
($t=710\,H_{\rf}/v_{\conv}$).  It is evident that strong magnetic fields are 
concentrated in the narrow, low-entropy vortical downdrafts characteristic 
of stratified convection (particularly in the upper half of the box), 
and that a greater proportion of the total unsigned magnetic flux 
resides in the lower half of the box.  The net signed flux in our dynamo 
simulations is always zero, and unless otherwise stated, all magnetic fluxes 
discussed hereafter refer to unsigned magnetic fluxes.  We note that even 
though our simulations exhibit the vertical flow asymmetries typical of 
stratified convection, and even though we use different values of 
dimensionless fluid parameters and do not achieve the highest numerical 
resolutions explored by \citet{Cattaneo99}, we still find similar and 
significant levels magnetic energy relative to kinetic energy over the 
simulation domain.  Figure~\ref{fig:mgrams} shows a ``magnetogram'' of the 
vertical component of the field near the top of the simulation box 
(left panel) and near the base of the box (right panel).
  
\begin{figure}
\epsscale{1.00}
\plotone{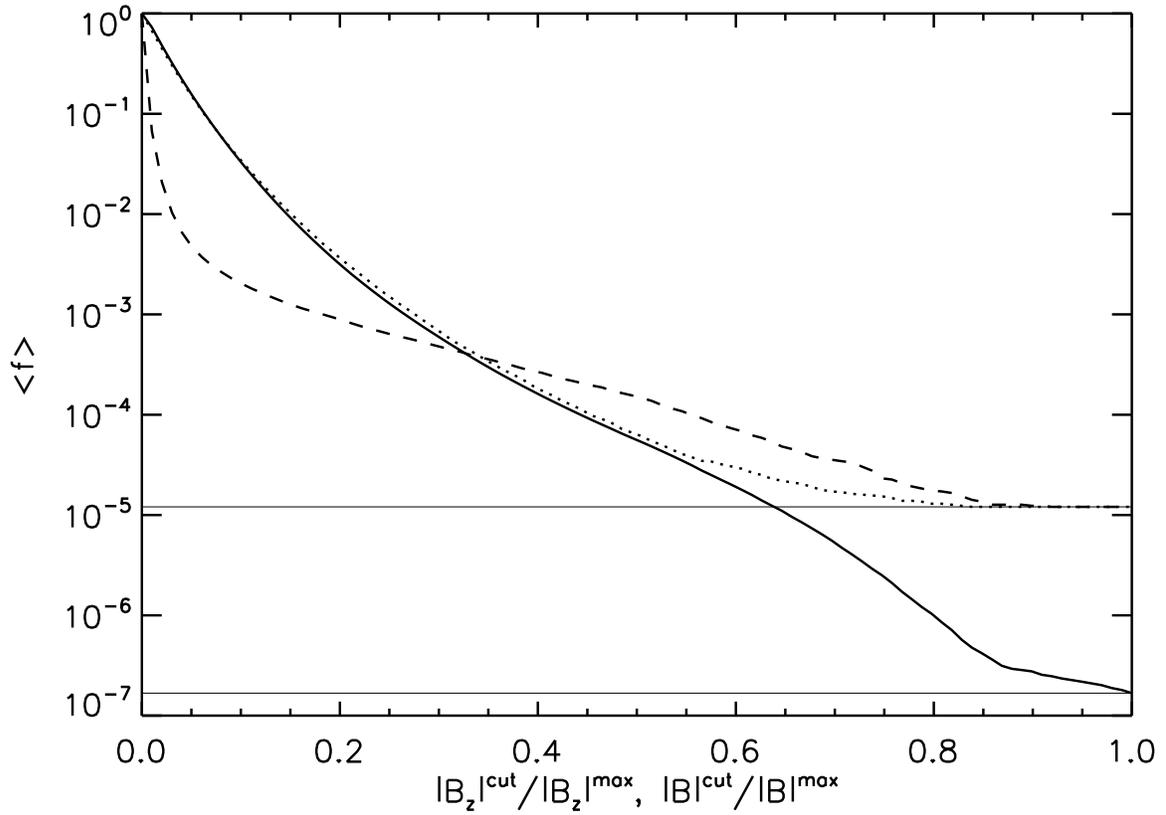}
\caption{The time-averaged filling factor $<f>$ as a function of 
$|B_z|^{\mathrm{cut}}/|B_z|^{\mathrm{max}}$
for two horizontal slices through the domain: 
near the top (dashed line), and near the bottom (dotted line).  Also shown 
is the volumetric filling factor over the entire domain (solid line).  The 
thin straight lines represent the one count per area and one count
per volume baselines.  See text for details.
\label{fig:filling_factors}}
\end{figure}

To quantify the distribution of unsigned magnetic flux in the domain
(after the dynamo has fully saturated), we define a magnetic filling factor 
\begin{equation}
   f=\frac{\int_0^{\ls{y}}\int_0^{\ls{x}} N dxdy}{\int_0^{\ls{y}}\int_0^{\ls{x}}dxdy}\, ,
\end{equation}
where $\ls{x}$ and $\ls{y}$ refer to the horizontal extent of the Cartesian 
domain,
and the quantity $N$ is defined as unity if $|B_z|\ge |B_z|^{\mathrm{cut}}$ 
and zero otherwise ($|B_z|^{\mathrm{cut}}$ is a chosen threshold value).  
If we adopt a threshold value of 
$|B_z|^{\mathrm{cut}}/|B_z|^{\mathrm{max}}=0.5$, we find that along 
horizontal slices near the top and bottom of the domain --- 
$z_u=2.06\,H_{\rf}$ and $z_l=0.45\,H_{\rf}$ respectively --- the time average
of the filling factor $<f>$ varies little, and is quite small:
$<f(z_u)>=1.47\times 10^{-4}$ and $<f(z_l)>=6.09\times 10^{-5}$. 
This indicates that strong fields are concentrated and highly localized.
Here, $|B_z|^{\mathrm{max}}$ refers to the maximum value of $|B_z|$ 
at $z_u$ sampled every $5\,H_{\rf}/v_{\conv}$ in the time 
interval between $t=600$--$850\,H_{\rf}/v_{\conv}$.
If we choose a cutoff of $|B_z|^{\mathrm{cut}}/|B_z|^{\mathrm{max}}=0.05$ 
we find a much larger disparity for the same horizontal planes: 
$<f(z_u)>=0.00473$ near the surface, and $<f(z_l)>=0.149$ near the bottom.  
This suggests that the weaker field is more evenly distributed, 
particularly in the lower half of the domain.  
Figure~\ref{fig:filling_factors} shows the time averaged filling factor 
as a function of $|B_z|^{\mathrm{cut}}/|B_z|^{\mathrm{max}}$ along both 
slices.  The dashed and dotted lines denote the horizontal planes near 
the surface and closer to the lower boundary respectively (these planes
are the same as those shown in Figure~\ref{fig:mgrams}).  The solid 
line denotes the time averaged volumetric filling factor (the volumetric 
analogue to the area filling factor with $|B_z|^{\mathrm{max}}$ defined 
as the maximum value of $|B|$ in the volume). 

A topic of great current interest is the dependence of dynamo behavior on
the magnetic Prandtl number $P_m$, which is the ratio of kinematic viscosity to
magnetic diffusivity, (or the ratio of the magnetic Reynolds 
number to the viscous Reynolds number, $R_m/R_e$).  Considering only the 
processes of Coulomb collisions in a fully ionized upper
stellar convection zone, one can show 
$P_m \sim 1.3 \times 10^{-5} T^4 / n$, where $T$ is the temperature in K, and
$n$ is the number density of hydrogen atoms in $\rm {cm}^{-3}$.  Substituting
typical values for $T$ and $n$, one finds $P_m << 1$.

While a complete investigation of the dependence of our own convective
dynamo on $P_m$
is beyond the scope of this paper, as the simulation described above took
5--6 CPU months of computing resources, we have performed several exploratory
calculations to get a rough idea of how the convective
dynamo effectiveness depends on
$P_m$.  First, we have used the initial convective state of the dynamo
run of this paper to explore how $P_m$ affects the early growth
phase of our convective dynamo.  The result is that the growth rate decreases
rapidly for $P_m \lesssim 2/3$, but shows signs of saturation
for $P_m \gtrsim 2$.  Second, we have used the saturated dynamo state of this
paper as the initial condition for a series of simulations in which the same 
viscous Reynolds number is used, but where the magnetic Reynolds number has 
been changed, resulting in
differing values of $P_m$.  The convective dynamo then begins to relax to
a new state that reflects the changed value of $P_m$.  
It appears that both the magnetic energy and the unsigned magnetic flux 
drop significantly for 
$P_m \lesssim 2/3$, but shows signs of saturation for $P_m \gtrsim 2$.  
Beyond this, for a fixed value of $P_m$,
we also find evidence for a dependence on the magnetic or viscous Reynolds 
number due to the fact that we cannot simulate the full inertial range.  We 
expect, as per the discussion below, that there will be an asymptotic limit 
to the kinetic and magnetic energies as the Reynolds numbers 
increase.

From this behavior, we conclude that for values of $P_m$ much smaller than
$2/3$ (for the range of magnetic and viscous Reynolds numbers that we can 
simulate), the convective dynamo is ineffective at generating magnetic
energies that approach those of the kinetic energy of convection.  We also 
conclude that values of $P_m \sim 1 - 2$ can generate significant amounts
of magnetic energy and flux, with levels that are within a factor of $\sim 10$
of equipartition with kinetic energy.  We speculate that larger values of
$P_m$ may to some degree increase the effectiveness of the dynamo beyond 
the cases we can investigate here,
but it seems unlikely that the dynamo can generate more
magnetic energy than the kinetic energy equipartition value.
Further, the simulations of \citet{Longcope03} , in which a closed 
magnetic ring is stretched
by random convective motions, essentially has $P_m = \infty$ since the
magnetic resistivity is zero in calculations like these.  Yet these
simulations still attain a finite amount of magnetic energy.  
In the large $P_m$ limit, the field strength is  most likely determined 
by a dynamic force balance between
ram pressure gradients and magnetic tension forces, rather than by a balance 
between field amplification via stretching and field decay by magnetic
diffusion, as occurs in the kinematic limit in many dynamo simulations.

Given the dependence of the convective dynamo on $P_m$ that we seem to
find, the small value of $P_m$ set by collisional processes in stellar 
convections zones suggests that convective dynamos are extremely inefficient.
However, a number of observational and theoretical considerations suggest
the appropriate value for $P_m$ is in fact much closer to unity in real
stellar convection zones.  We discuss these here.

First, magnetic elements on the Sun are observed to move, disperse, and
cancel in response to granular and supergranular motions 
\citep{Simon01,Parnell02}, indicating that long before the 
tiny molecular dissipation scales are reached, motions associated with
convection can effectively disperse a magnetic element on much larger size 
scales.  This means that the effective magnetic diffusivity is much 
larger than the molecular value.  A similar argument applied to the observed
correlation length and correlation time
of convective motions also leads to a kinematic viscosity
much greater than the molecular value, and qualitatively close to the
effective magnetic diffusivity -- implying a value of $P_m$ near unity.

A related theoretical argument for $P_m \sim 1$ comes from two 
observations of our own simulations: 
(1) Both the magnetic and viscous Reynolds numbers of the
real physical system we are attempting to model are far greater than those
in our code.  Therefore, at the smallest scale we can resolve in our code,
magnetic fields and momentum are still well within the inertial range
of the turbulent motions excited by convection, and therefore
both momentum and magnetic fields will be diffused by eddies
at the resolution scale and smaller.  Applying a common eddy diffusivity
for both quantities leads to a value of $P_m \sim 1$.  A pitfall to this
argument would occur if one could show that what occurred on 
sub-resolution scales
could strongly affect what happens on the largest scales.  This leads
to our second observation:  (2) Both the magnetic and kinetic energy spectra
exhibit peaks at large scales, with significantly reduced energy at the smallest
scales in the simulation, especially near the bottom of the simulation where
the magnetic energy density is highest.
Given this, it seems unlikely that details of the dissipation physics occurring
on unresolved scales will have a strong effect on the macroscopic properties of
the dynamo simulation we have computed using $P_m = 4/3$.  It is interesting
to note that our conclusion that both magnetic and kinetic energies peak at
large scales differs from the results of dynamo calculations using the
``Kazantsev'' approach, in which either the 
velocity (in kinematic models, e.g.  \citealt{Boldyrev04})
or an external force field (in dynamic models, e.g.  \citealt{Schekochihin04})
are driven by an assumed temporal white noise spectrum.
Because our driving motions are computed self-consistently from the convection
itself, and not from ad-hoc assumptions about a forcing term,
we feel we have applied our results in a physically self-consistent manner
by assuming a value of $P_m$ near unity.

There are other theoretical arguments against using the molecular diffusivities.
\citet{Longcope03} argue that the presence of fibril magnetic
fields in the convection zone plasma 
results in an effective viscosity that is at least as
great as the turbulent viscosity, if not larger.  \citet{Abbett04}
show that magnetic flux tubes are
dispersed in a model convection zone with an effective magnetic diffusivity 
given by a turbulent eddy diffusivity.

Taken together, all of these results
argue for a common eddy diffusivity for both 
processes and therefore a magnetic Prandtl number near unity.
We adopt $P_m = 4/3$ instead of unity simply because we want to choose
the largest possible values of the viscous and magnetic Reynolds numbers
that do not result in numerical artifacts.
Our preliminary calculations indicate no
significant differences between using results from 
$P_m = 4/3$ versus assuming a value of unity.

\section{Connecting the Dynamo Model to the Stellar Envelope}\label{sec:connect}
Here, we relate the anelastic simulation results described in 
\S \ref{sec:description}, which are performed in dimensionless units, to 
physical units so that the resulting magnetic surface flux 
can be compared with 
stellar observations.  We must therefore assign cgs units to the 
non-dimensional simulation results and connect the velocity and magnetic 
fields of the simulation to the luminosity, surface gravity, radius, effective
temperature and surface density of main sequence reference stars.

The variation of the background temperature $T_0(z)$ and density $\rho_0(z)$
with height $z$ above the base of our anelastic simulation box is taken to be
a polytrope with index $m=1/(\gamma-1)=1.5$ (here $\gamma$ refers to the adiabatic
index for an ideal gas) and is given explicitly by 
\begin{equation}
  T_0(z)=T_{\rf}\left[1-\frac{z}{(m+1)H_{\rf}}\right]
  \label{eq:tzero}
\end{equation}
and
\begin{equation}
  \rho_0(z)=\rho_{\rf}\left(\frac{T_0(z)}{T_{\rf}}\right)^m\, .
\label{eq:rhozero}
\end{equation}
Here, the subscript ``ref'' denotes chosen reference values at the base of 
the box (see \citealt{Fan03}).  Values of the thermodynamic variables 
at the top (or surface) of the box are denoted with a subscript ``surf''.  
The ratio $T_{\rf}/T_{\surf}$ is one of the fixed parameters in the 
simulation described in this paper, and is set to $7.8$, and results in a 
stratified background atmosphere with a base-to-surface density ratio 
$\rho_{\rf}/\rho_{\surf}$ of $21.8$ and a corresponding pressure ratio 
$P_{\rf}/P_{\surf}$ of $170.6$.  The reference temperature $T_{\rf}$ at the 
base is determined from the effective surface temperature $T_{\mathrm{eff}}$ 
by setting $T_{\surf}=T_{\mathrm{eff}}$, and then using the fixed ratio 
$T_{\rf}/T_{\surf}$ to determine the actual value of $T_{\rf}$ in degrees K.  
Similarly, the value of surface density from stellar atmosphere models, along
with equation \ref{eq:rhozero} is used to determine the reference density 
$\rho_{\rf}$.
                                                                                    
In the anelastic simulation, the unit of length is the pressure 
scale height at the base of the domain, 
$H_{\rf}\equiv R T_{\rf}/(\bar{\mu} g_{\rf})$, where 
$R$ is the universal gas constant, $\bar{\mu}$ is the mean molecular weight, 
and $g_{\rf}$ is the value of the gravitational acceleration at the base of 
the simulation box.  For simplicity, we assume that $\bar{\mu}=0.5$, which 
corresponds to a fully ionized hydrogen gas.  Near the surface of the later 
type stars we consider, the gas will be nearly neutral, however.  Our neglect 
of the effects of variable ionization in the background atmosphere is a 
limitation of our current calculations that could be improved in future work. 
We assume that the values of $g_{\rf}$ and $g_{\surf}$, 
the surface gravitational acceleration, are equal, since the depth
of the box is quite small compared to a stellar radius for the cases we 
consider.  The depth of the box $\ls{z}$ is given by 
\begin{equation}
  \ls{z}=(m+1)\frac{T_{\rf}-T_{\surf}}{T_{\rf}}H_{\rf} \, ,
  \label{eq:lz}
\end{equation}
and the horizontal dimensions of the box, $\ls{x}$ and $\ls{y}$ are four 
times larger than $\ls{z}$.

The unit of velocity in the simulations is defined in terms of a 
dimensionless entropy gradient (see \S~\ref{sec:description}).  To convert 
the velocity units to physical values, we note that in our simulation, the 
dominant form of energy transport from the base to the top is thermal 
convection.  We therefore require that the vertical energy flux match 
that needed to carry the stellar luminosity in a particular stellar model.  
In other words, for a given star with luminosity $L_{\star}$ and radius 
$R_{\star}$, the energy flux carried by convection in the outer convective 
envelope must be given by
\begin{equation}
  F_{\conv}=\frac{L_{\star}}{4\pi R_{\star}^2} .
  \label{eq:fconv}
\end{equation}
We have assumed that throughout the convective envelope, the fraction
of energy carried by radiative diffusion is far less than that carried by
convection, which is a reasonable assumption for the stars we consider
here.  This assumption will break down near the photosphere, where radiation
begins to dominate.

To estimate $F_{\conv}$ in a consistent way that applies
to each of the different stellar types we consider, we use a simple
form of mixing length theory to obtain an approximate reference value of 
the convective velocity at the depth corresponding to the base of our 
simulation domain.  Here, we adopt the mixing length formulation of 
\citet{Mihalas78}.  Combining his equations 7-68 and 7-69, and explicitly 
evaluating $C_p$ as $5R/(2\bar{\mu})$, we find
\begin{equation}
  F_{\conv}=\frac{10}{\alpha}\rho v_{\conv}^3 \, ,
  \label{eq:mlt}
\end{equation}
where $\rho$ is the background density of the convective envelope, $v_{\conv}$
is the characteristic convective velocity, and $\alpha=l/H_p$
is the ratio of the ``mixing length'' to the pressure scale height.  While
mixing length theory is a crude approximation, it has been shown 
to provide roughly the right velocity amplitude in simulations of realistic 
surface convection \citep{Abbett97}, who also find best results when 
$\alpha\sim 1.5$.  Our use of equation \ref{eq:mlt} is to take the stellar 
value of $F_{\conv}$ and the background stratification of $\rho$ to 
estimate the value of $v_{\conv}$ at the base of our simulation (which we will 
take as the reference value $v_{\rf}$), and to then assume that our simulation 
velocity is measured in units of this value.  If desired, one can work 
backwards to determine the corresponding entropy gradient, but this is not 
necessary.  We have used equations \ref{eq:fconv} and \ref{eq:mlt}
to estimate the convective velocity in the layers just below the solar 
photosphere and find values of $\sim 3$ km $\mathrm{s}^{-1}$, in 
approximate agreement with much more detailed models of the solar interior 
(see e.g., \citealt{Abbett97,Stein98,Asplund00,Samadi03}), and with 
observations of convective velocities from granulation 
\citep{Hirzberger01,Roudier03}. 
                                                                                    
Because the depth of our simulation box is small compared to the 
stellar radius in all the cases we consider, we are justified in
ignoring the change in radius with depth and assume that $F_{\conv}$
is uniform with depth within the box.  This means that $v_{\conv}$ at the 
base of the box (i.e., $v_{\rf}$) can be found from equations \ref{eq:tzero}, 
\ref{eq:rhozero}, \ref{eq:fconv}, and \ref{eq:mlt}, and is given by 
\begin{equation}
  v_{\rf}=v_{\surf}\left(\frac{T_{\surf}}{T_{\rf}}\right)^{\frac{m}{3}} \, ,
  \label{eq:vr}
\end{equation}
where 
\begin{equation}
  v_{\surf}=\left(\frac{\alpha L_{\star}}{40\pi\rho_{\surf}R_{\star}^2}\right)^{\frac{1}{3}} \, . 
  \label{eq:vs}
\end{equation}
Note that equation \ref{eq:vr} is extremely insensitive to the precise
formulation used in the mixing length theory.  For example, if the 
coefficient ``$10$'' in equation \ref{eq:mlt} was changed by a factor of ten, 
it would result in only a factor of $2.15$ difference in the resulting value 
of $v_{\rf}$.

The magnetic field strengths of the simulation are given in units of
\begin{equation}
  B_{\rf}=(4\pi\rho_{\rf})^{\frac{1}{2}} v_{\rf} \, .
  \label{eq:bunit}
\end{equation}
To convert to physical units (G), we simply substitute 
the expression for $v_{\rf}$ (equation~\ref{eq:vr}) into equation~\ref{eq:bunit}:
\begin{equation}
  B_{\rf}=\left[\frac{\pi^{\frac{1}{2}} \alpha L_{\star}}{5 R_{\star}^2}\right]^{\frac{1}{3}} 
    \rho_{\surf}^{\frac{1}{6}} \left[ \frac{T_{\rf}}{T_{\surf}}\right]^{\frac{m}{6}} \, .
  \label{eq:br} 
\end{equation}

The scaling relationships given here allow us to estimate the amount of 
magnetic flux and the mean field strength near the top of our anelastic
simulation of the turbulent dynamo.  We note that although the anelastic
formulation is not well suited to modeling the surface layers of stellar 
atmospheres, and that the convection in our simulations is not driven 
explicitly by radiative cooling in the surface layers, our simulation remains 
appropriate for our study since we are only interested in obtaining an 
estimate of the total magnetic flux threading the top of the convective 
envelope --- not the detailed dynamics or distribution of magnetic fields 
across a stellar photosphere. 
Table~\ref{tbl:stellar_params} lists the parameters used to calculate
the scaling of the simulation results for a sample of main sequence 
stars ranging in spectral type from F0 to M0.  All stellar parameters except 
surface density are taken from \citet{Gray92} for spectral types F0--K5, 
and those for M0 are taken from \citet{Reid00}.  The surface densities 
(column $6$ of Table~\ref{tbl:stellar_params}) are evaluated at the depth
where the local temperature is equal to the effective temperature 
(column $2$) using model atmospheres from \citet{Kurucz93}.

\section{Results}\label{sec:results}
We have generated significant magnetic fields (via a convective dynamo) in 
a stratified turbulent model convection zone by imposing a dynamically 
insignificant seed field on a statistically relaxed convective state. 
Our treatment differs from that of \citet{Cattaneo99}, since our domain 
is highly stratified; however, the energetics of our simulation are 
similar to his results --- we find that the magnetic field fully 
saturates at roughly seven percent of the total kinetic energy of the 
computational domain.  We find that the magnetic filling factor is small 
near the surface, and larger deeper in the convective envelope, and that 
the total amount of unsigned magnetic flux is concentrated in the lower 
half of the simulation domain.  Strong fields are concentrated and highly 
localized, while weaker fields are more evenly distributed.  We also note 
that the correlation time of magnetic flux structures near the top of the 
simulation box is $2$--$3\,H_{\rf}/v_{\conv}$, which corresponds to 
$1$--$2$ hours for a solar-type star.  

The results of applying the $\alpha=1.5$ mixing length scaling to the 
simulation data are presented in columns $2$--$4$ of 
Table~\ref{tbl:scaled_results}: column $2$ lists the pressure scale
height at the base of the domain, column $3$ lists the convective velocities 
at the surface, and column $4$ lists the total magnetic fluxes of the 
reference stars, $\Phi_{\star}$.  The magnetic fluxes were obtained by 
first averaging the total unsigned flux through a horizontal layer near 
the top of the simulation box ($z=z_{u}$) 
over an interval $250\,H_{\rf}/v_{\conv}$ 
centered at $t=725\,H_{\rf}/v_{\conv}$ 
(see Figure~\ref{fig:log_energy_relaxation}).  
The simulated result was then scaled from the horizontal area of the 
computational domain to the surface area of the reference stars.  Here, 
we assume that all areas on the stellar surface are uncorrelated and 
generate magnetic field in the same way and with the same efficiency.  

We estimate the X-ray luminosities of the reference stars through an 
empirical relationship between the X-ray luminosity and unsigned magnetic flux 
\citep{Pevtsov03}.  A fit to the data presented in Figure $1$ of that paper
leads to the relation
\begin{equation}
  L_{\xray}=0.8940 \Phi_{\star}^{1.1488} \, .
\end{equation}
The X-ray luminosity, $L_{\xray}$, and surface X-ray flux, $F_{\xray}$, 
for our sample of reference stars are listed columns $4$ and $5$ of 
Table~\ref{tbl:scaled_results}, respectively.

We compare the results of our simulation to the volume limited sample
of cool stars discussed in \citet{Schmitt97}.  The stellar data itself
comes from both \citet{Schmitt95} and \citet{Schmitt97}.  These papers
provide the X-ray luminosity and either the absolute visual magnitude,
$M_V$, or the spectral type and $M_V$ for each star.  The observed
$B - V$ is also given for about half the stars in the sample.  The
spectral type (if given) or $M_V$ is converted to a $B - V$ index (if
needed) and a stellar radius using the main sequence calibration given
by \citet{Gray92} for stars K7 and earlier, or that given by \citet{Reid00}
for stars M0 and later.  In the case of the M dwarfs, \citet{Reid00} 
do not give a direct spectral type --- $M_V$ (or $B - V$) calibration, 
so we used the data provided in their appendices to construct
one.  We binned the stellar data for the $8$ pc sample into spectral type
bins (M0, M0.5, M1, M1.5, \dots) and averaged the values of $M_V$ and
$B - V$ in each bin.  These averages were then fit with a 5th-order
polynomial to construct our final M dwarf calibration of $M_V$ and $B - V$
as a function of spectral type.  Using the published X-ray luminosities
and our calculated stellar radii, we then found the X-ray surface flux for
each observed star (shown as a filled circle in Figure~\ref{fig:xrayflux}).  
Our predicted model surface fluxes calculated using a
mixing length parameter $\alpha=1.5$ are shown with asterisks connected by a 
dashed line in the same figure.  The gray region shows the range of the 
predicted surface X-ray fluxes if the convective flux calculated by the 
models varies by a factor of eight (or equivalently if the velocity
scaling varies by a factor of two) above or below that
calculated using $\alpha=1.5$. 

It is reasonable to assume that the heating mechanism that produces 
X-ray emission also produces emission in chromospheric and transition
region diagnostics, contributing to the basal emission level
observed in these lines.  Without a detailed magnetic heating model, we
can only explore this issue using flux-flux relationships between X-ray 
emission and diagnostics such as \ion{Mg}{2} emission.  Many investigators 
have found good correlation between these diagnostics, and here we use the 
flux-flux relationship
\begin{equation}
  \log{F_{\mg}}=\frac{\log{F_{\xray}}}{1.97}+\frac{6.7}{1.97}
  \label{eq:fluxflux}
\end{equation}
based on Table 4 of \citet{Rutten91} to determine an estimate of 
\ion{Mg}{2} emission for our reference stars.  

\citet{Rutten91} do not find a basal emission level in their X-ray
observations, but do find one in the \ion{Mg}{2} emission which they attribute
to acoustic heating.  Subtracting off this basal emission from observed
\ion{Mg}{2} emission for all stars then improves the correlation of \ion{Mg}{2}
with X-ray emission.  The assumption is that this excess \ion{Mg}{2} emission
is all magnetic in origin, as is the X-ray emission.  Since the emission we 
calculate is explicitly magnetic in origin, we use the relationship between 
excess \ion{Mg}{2} emission and X-ray emission (column b of Table 4 in 
\citealt{Rutten91}).  We note though, that using the total flux-flux 
relationship given in \citet{Rutten91} increases the predicted \ion{Mg}{2} 
emission by only $\sim 0.2$ dex, which is well within the range of variation 
shown in the plots for different choices of the mixing length in our 
magnetoconvection simulations.  Since our predicted X-ray surface fluxes do 
not depend strongly on stellar $B - V$, it is not surprising that we find a 
similarly weak dependence on stellar spectral type of the resulting \ion{Mg}{2}
surface flux: we find $\log (F_{\mg}) = 5.50$ at F0 and 
$5.34$ at M0.
  
At first glance, these values appear inconsistent with the basal flux
observations of \citet{Rutten91} (their figure 1b), where the basal level
appears to be $\log(F_{\mg}) \sim 4.5$ at $B - V = 1.5$.  However, a 
closer inspection of this figure (indeed, of all the flux-color figures in
\citet{Rutten91}) shows that the lower limit to the \ion{Mg}{2} emission is 
defined by giants and subgiants at later spectral types.  The dwarf stars are
all significantly stronger than the apparent lower limit at these later 
spectral types.  In addition, there is no apparent piling up of the dwarf 
star observations against a lower bound as would be expected for basal 
emission (see \citealt{Schrijver95}), probably due in part to the small 
number of dwarf stars observed by \citet{Rutten91} (there are only 13 stars 
later than $B - V = 1.0$ in their Figure 1b).  This lack of piling up of the 
dwarfs is apparent across all $B - V$ (0.3 to 1.5) plotted in 
\citet{Rutten91}.  A more complete survey of \ion{Mg}{2} emission in K and M 
dwarf stars is provided by \citet{Mathioudakis92}.  In 
Figure~\ref{fig:mg2flux} we plot the \ion{Mg}{2} surface flux on the dK and 
dM stars from Tables 1 and 2 of \citet{Mathioudakis92} against their observed 
$B - V$ (we do not plot the dKe and dMe stars as these stars are expected to 
--- and do --- show much stronger emission than that from basal emission 
stars).  Also shown in Figure~\ref{fig:mg2flux} is our predicted \ion{Mg}{2} 
surface flux based on our dwarf models from equation~\ref{eq:fluxflux}.  
Our predicted emission levels due 
to a convective dynamo fall nicely at the lower limit of the observed 
emission in the range that the models are valid ($B - V < 1.5$).  The 
agreement seen in Figure~\ref{fig:mg2flux} is substantially better than that 
shown in the acoustic heating models of \citet{Buchholz98}, whose predicted 
surface fluxes are too weak ($\log(F_{\mg}) = 4.6$ to $4.8$ at 
$B - V = 1.4$) and show a much stronger color dependence (for $B - V < 1.4$) 
than seen in the dwarf star observations of \citet{Mathioudakis92} or seen 
in the dwarf star observations of \cite{Rutten91}.  On the other hand, at 
earlier spectra types ($B - V \lesssim 0.5$) the acoustic heating models better 
match the lower bound of the \ion{Mg}{2} observations than the crude estimates 
presented here.

As discussed in \S~\ref{sec:description}, the overall flux achieved in 
our convective dynamo simulations does depend on the value of the magnetic
Prandtl number $P_m$.  However, the
scaling of the model to main sequence stars of different spectral types
does not---the resulting weak color dependence of the X-ray
and \ion{Mg}{2} emission is independent of $P_m$, whereas 
acoustic heating models show a strong color dependence.  The
X-ray data (Figure~\ref{fig:xrayflux}) and the more complete \ion{Mg}{2} 
data (Figure~\ref{fig:mg2flux})
show a weak color dependence over the spectral type range in our models
that is consistent with the convective dynamo predictions and not with
the acoustic model predictions.

\section{Discussion and Conclusions}\label{sec:discussion}
\begin{figure}
\epsscale{1.00}
\plotone{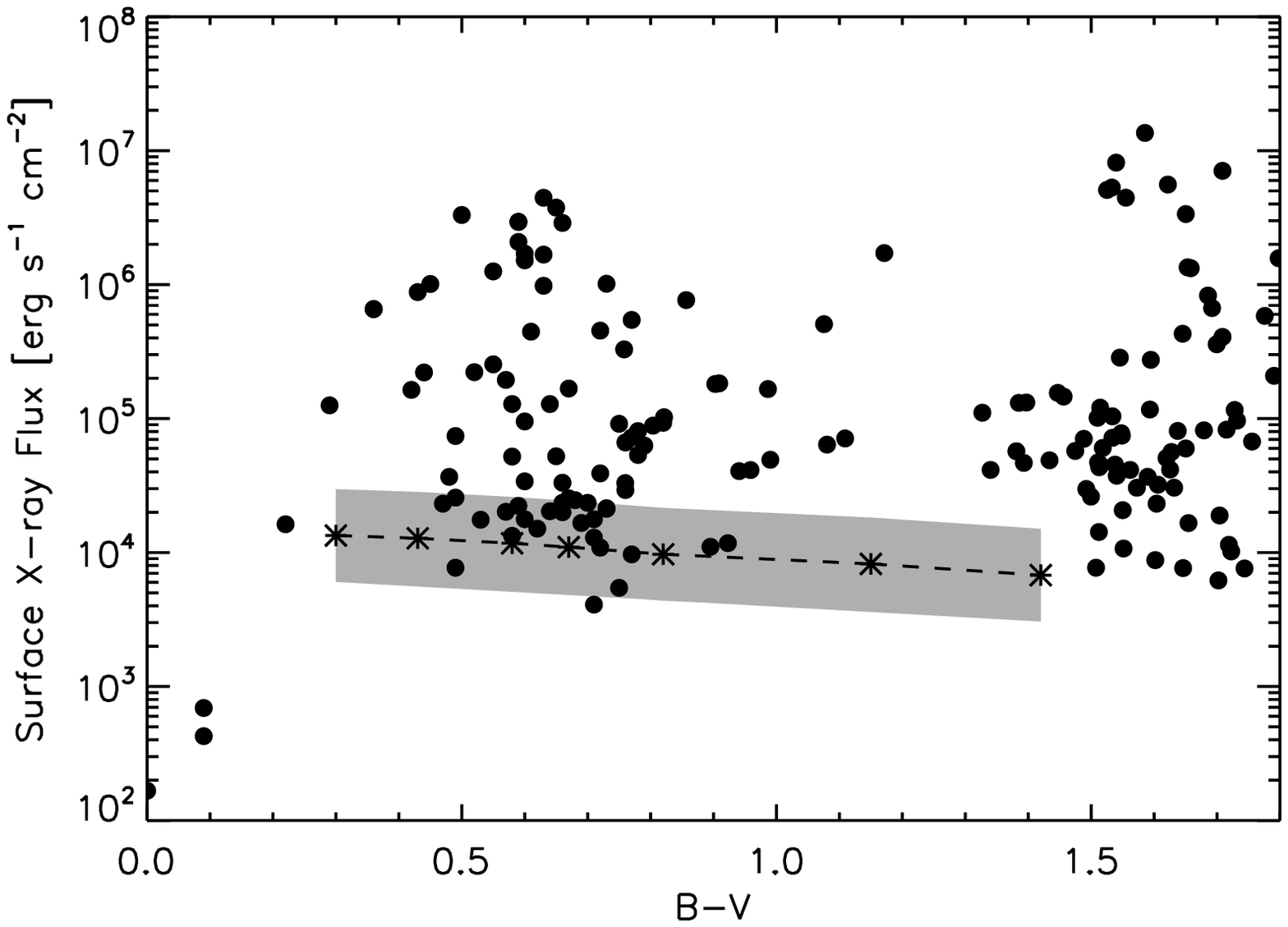}
\caption{The filled circles represent the X-ray surface flux for
each observed star as a function of $B - V$ color.  The asterisks
represent our theoretical prediction of the lower bound of the
X-ray surface flux for a choice of mixing length parameter
$\alpha=1.5$.  The gray shaded area indicates the amount
that these levels can change if the assumed surface velocities
change by a factor of two.  See text for details.}
\label{fig:xrayflux}
\end{figure}

We have performed 3D MHD simulations of a turbulent dynamo in
a highly stratified Cartesian domain in order to determine 
the amount and distribution of magnetic flux generated in a
non-rotating convective envelope.  We use the computed
surface flux near the upper boundary of the domain
and an empirical relationship between magnetic flux and X-ray 
flux \citep{Pevtsov03} to determine the lower limit of X-ray 
emission in main sequence stars.  Figure~\ref{fig:xrayflux} 
suggests that our simple analytic scaling treatment successfully 
reproduces the observed lower limit of X-ray flux found by 
\citet{Schmitt97} in the range from F0--M0.  This result suggests
that the level of heating by magnetic sources in the coronae of these
stars is sufficient to account for most if not all of the X-ray flux.  

Stars earlier than spectral type F0 are expected to have the 
thicknesses of their outer convective shells rapidly decrease to zero as 
the surface temperature increases.  The low surface X-ray fluxes for the 
three observed stars for which $B - V < 0.1$ are consistent with less 
heating due to a diminished level of magnetic field generated by the 
convective dynamo;  however, the same argument holds for less heating due 
to a diminished convective acoustic flux.  We do not extend our treatment 
to spectral types later than M0, since we expect that our assumptions of 
an ideal equation of state and high electrical conductivity will 
break down for cool M dwarf atmospheres.  Our treatment implicitly assumes 
that the \citet{Pevtsov03} relation applies to magnetic flux generated by 
the turbulent convective dynamo.  Currently, the best hope for verifying this 
assumption is forward modeling of coronal heating of ensembles of magnetic 
loops (e.g., \citealt{Lundquist04,Schrijver04}), as well as in 
understanding the role of sub-pixel magnetic structures on measured 
magnetic flux densities. 

\begin{figure}
\epsscale{1.00}
\plotone{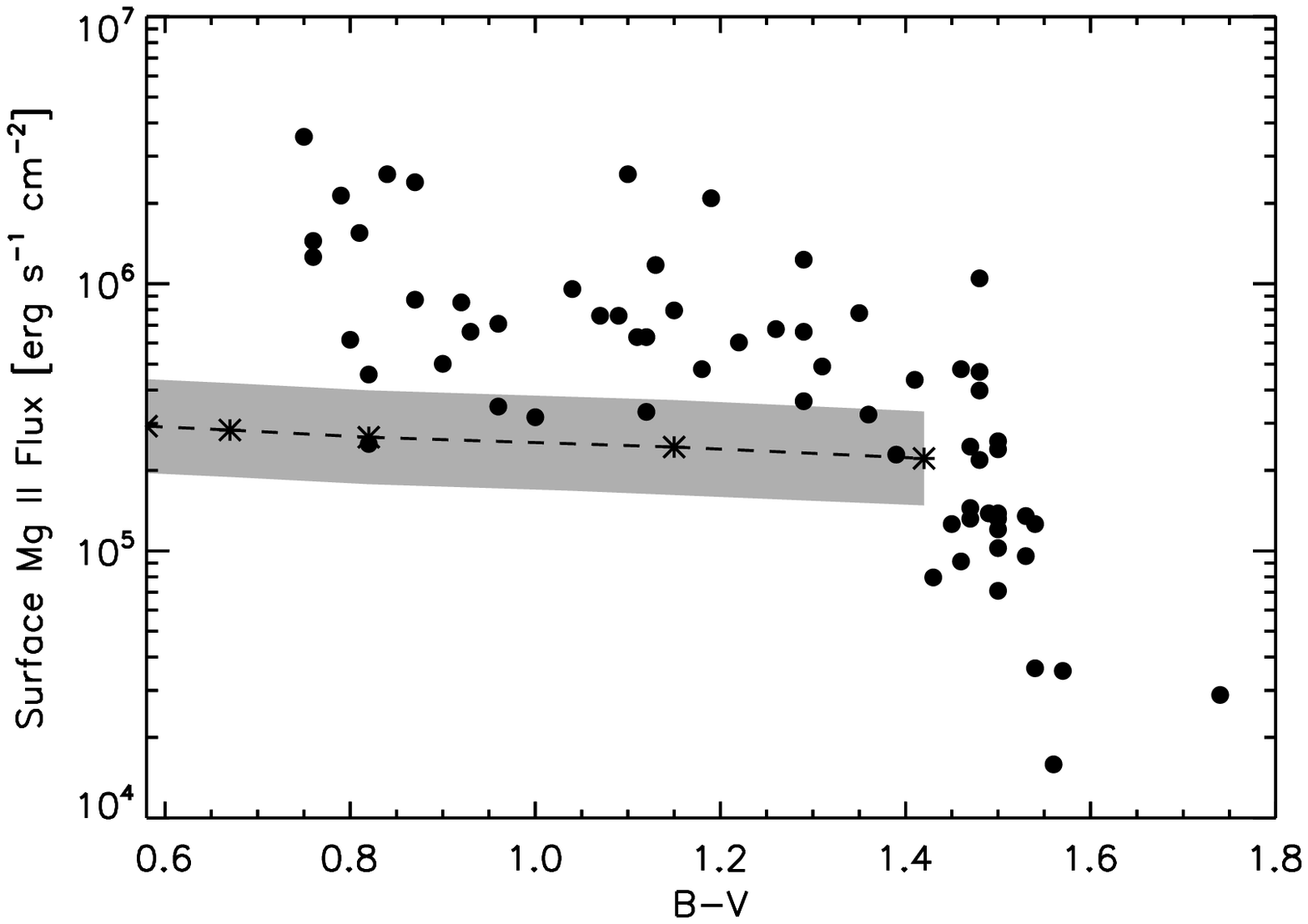}
\caption{The filled circles represent the \ion{Mg}{2} surface flux for
each observed star as a function of $B - V$ color.  The asterisks
represent our theoretical prediction of the lower bound of the
\ion{Mg}{2} surface flux for a choice of mixing length parameter
$\alpha=1.5$.  The gray shaded area indicates the amount
that these levels can change if the assumed surface velocities
change by a factor of two.  See text for details.}
\label{fig:mg2flux}
\end{figure}

We use the flux-flux relation of \cite{Rutten91} to
determine the expected level of chromospheric emission in \ion{Mg}{2}.
For K dwarfs we find good agreement with the observed lower limit of
\ion{Mg}{2} surface flux in the K dwarf sample of \citet{Mathioudakis92},
suggesting that this observed lower limit is entirely the result of
magnetic heating.  On the other hand, our models predict a very shallow
dependence of both X-ray and \ion{Mg}{2} emission with spectral type, so that
they underestimate the \ion{Mg}{2} emission at earlier spectral types
(though our models are a good match to the X-ray emission at these
spectral types).  This suggests some other agent (acoustic heating) is
required to produce the minimum observed \ion{Mg}{2} emission at earlier
spectral types, and that this additional mechanism produces relatively
little X-ray emission.  At the very latest spectral types, the sharp drop
in the \ion{Mg}{2} surface flux seen in Figure~\ref{fig:mg2flux} at 
$B - V \sim 1.5$ suggests something very different may be happening here.
We do not expect our models to be valid at these cool temperatures due to
the relatively low ionization that will result.  In addition, the X-ray
emission from stars observed in Figure~\ref{fig:xrayflux} does not show 
this same drop, indicating that the flux-flux calibration used to predict
the \ion{Mg}{2} emission based on the X-ray emission will not hold.  We 
cannot address the level of acoustic heating with our dynamo model.  Obviously, 
if there is convection there will be some level of acoustic activity.
However, the results we present here provide a consistent and viable
alternative to acoustic heating for K type dwarfs in the absence of a 
large scale dynamo.  At earlier spectral types, our results suggest
turbulent dynamos can also fully account for the lower limit to the observed
X-ray emission, though there may still be a substantial acoustic 
contribution to the coronal emission in earlier type stars.

\acknowledgments
This work was supported by NASA's Astrophysics Theory Program through
the grant ``Three Dimensional MHD Simulations of Turbulent Dynamos in 
Convecting Stars''.  In addition, W.~P.~A. and G.~H.~F. were supported by 
NASA through the SEC SR\&T grant ``The Physics of Active Regions'', and 
through the DoD MURI grant, ``Understanding Magnetic Eruptions on the
Sun and their Interplanetary Consequences''.  C.~M.~J.-K. would like to 
acknowledge partial support from the NASA Origins program through grant 
number NAG5-13103. We wish to thank Gibor Basri, Yuhong Fan, Alex Pevtsov, 
Karel Schrijver, Bob Stein and the referee for helpful comments and 
discussions. 


\clearpage
\clearpage
\begin{deluxetable}{ccccccc}
\tabletypesize{\scriptsize}
\tablecaption{Stellar model parameters}
\tablewidth{0pt}
\tablehead{
\colhead{Spectral Type} & 
\colhead{$T_{\mathrm{eff}}$ [K]} & 
\colhead{$L_{\star}$ [erg $\mathrm{s}^{-1}$]} & 
\colhead{$R_{\star}$ [cm]} & 
\colhead{$\log g$ [cm $\mathrm{s}^{-2}$]} & 
\colhead{$\rho_{\surf}$ [g $\mathrm{cm}^{-3}$]}
}
\startdata
F0 & 6949 & $1.84\times10^{34}$ & $1.05\times10^{11}$ & 4.27 & 
    $6.71\times10^{-8}$ \\
F5 & 6445 & $1.04\times10^{34}$ & $9.19\times10^{10}$ & 4.32 &
    $1.18\times10^{-7}$ \\
G0 & 5948 & $5.04\times10^{33}$ & $7.52\times10^{10}$ & 4.42 &
    $1.97\times10^{-7}$ \\
G5 & 5678 & $3.31\times10^{33}$ & $6.68\times10^{10}$ & 4.46 &
    $2.42\times10^{-7}$ \\
K0 & 5273 & $1.75\times10^{33}$ & $5.64\times10^{10}$ & 4.53 &
    $3.02\times10^{-7}$ \\
K5 & 4557 & $6.88\times10^{32}$ & $4.73\times10^{10}$ & 4.60 &
    $5.33\times10^{-7}$ \\
M0 & 3800 & $2.77\times10^{32}$ & $4.32\times10^{10}$ & 4.65 & 
    $9.60\times10^{-7}$ \\
\enddata
\label{tbl:stellar_params}
\end{deluxetable}
\begin{deluxetable}{cccccccc}
\tabletypesize{\scriptsize}
\tablecaption{Scaled results for $\alpha=1.5$}
\tablewidth{0pt}
\tablehead{
\colhead{Spectral Type} & 
\colhead{$H_{\rf}$ [Mm]} &
\colhead{$v_{\surf}$ [km $\mathrm{s}^{-1}$]} &
\colhead{$\Phi_{\star}$ [Mx]} &
\colhead{$L_{\xray}$ [erg $\mathrm{s}^{-1}$]} &
\colhead{$\log F_{\xray}$ [erg $\mathrm{s}^{-1}$ $\mathrm{cm}^{-2}$]}
}
\startdata
F0 & $4.86$ & $6.67$ & $6.01\times10^{23}$ & $1.86\times10^{27}$ & $4.13$ \\
F5 & $4.02$ & $4.99$ & $4.57\times10^{23}$ & $1.36\times10^{27}$ & $4.11$ \\
G0 & $2.95$ & $3.78$ & $2.99\times10^{23}$ & $8.34\times10^{26}$ & $4.07$ \\
G5 & $2.56$ & $3.32$ & $2.30\times10^{23}$ & $6.16\times10^{26}$ & $4.04$ \\
K0 & $2.03$ & $2.79$ & $1.54\times10^{23}$ & $3.88\times10^{26}$ & $3.99$ \\
K5 & $1.49$ & $1.90$ & $9.81\times10^{22}$ & $2.31\times10^{26}$ & $3.92$ \\
M0 & $1.11$ & $1.23$ & $7.08\times10^{22}$ & $1.59\times10^{26}$ & $3.83$ \\
\enddata
\label{tbl:scaled_results}
\end{deluxetable}

\end{document}